\documentclass[aps,prb,twocolumn,superscriptaddress]{revtex4}
\usepackage{graphicx}
\usepackage{times}
\usepackage{amsmath}
\usepackage{textcomp}
\newcommand{\boldnabla}{\mbox{\boldmath$\nabla$}}

\begin{document}

\title{Optimizing the Majorana character of SQUIDs with topologically non-trivial barriers}

\author{M. Veldhorst}
\affiliation{Faculty of Science and Technology and MESA+ Institute for Nanotechnology, University of Twente, 7500 AE Enschede, The Netherlands}
\author{C.G. Molenaar}
\affiliation{Faculty of Science and Technology and MESA+ Institute for Nanotechnology, University of Twente, 7500 AE Enschede, The Netherlands}
\author{C.J.M. Verwijs}
\altaffiliation[Currently at ]{ASML, Veldhoven, The Netherlands}
\affiliation{Faculty of Science and Technology and MESA+ Institute for Nanotechnology, University of Twente, 7500 AE Enschede, The Netherlands}
\author{H. Hilgenkamp}
\altaffiliation[Also at ]{Leiden Institute of Physics, Leiden University, P.O. Box 9506, 2300 RA Leiden, The Netherlands}
\affiliation{Faculty of Science and Technology and MESA+ Institute for Nanotechnology, University of Twente, 7500 AE Enschede, The Netherlands}
\author{A. Brinkman}
\affiliation{Faculty of Science and Technology and MESA+ Institute for Nanotechnology, University of Twente, 7500 AE Enschede, The Netherlands}\date{\today}

\begin{abstract}
We have modeled SQUIDs with topologically non-trivial superconducting junctions and performed an optimization study on the Majorana fermion detection. We find that the SQUID parameters $\beta_L$, and $\beta_C$ can be used to increase the ratio of Majorana tunneling to standard Cooper pair tunneling by more than two orders of magnitude. Most importantly, we show that dc SQUIDs including topologically trivial components can still host strong signatures of the Majorana fermion. This paves the way towards the experimental verification of the theoretically predicted Majorana fermion.
\end{abstract}

\maketitle

\section{Introduction}
Superconducting junctions with topologically non-trivial barriers are predicted to host Majorana bound states \cite{Fu2008, Tanaka2009}. Non-trivial states include the edge or surface of the recently discovered topological insulators \cite{Zhang2006, Fu2007, Zhang2009N, Qi2009, Hsieh2008, Chen2009, Hsieh2009N, Peng2009, Cheng2010, Zhang2009P} and semiconducting nanowires in the presence of Rashba spin orbit coupling and a Zeeman field \cite{Sau2010, Alicea2010}. Candidates with high potential for the detection and manipulation of the Majorana fermion \cite{Majorana1937} are superconducting quantum interference devices (SQUIDs) \cite{Fu2009, Beenakker2011}. The appearance of Majorana bound states in superconducting junctions enables tunneling of quasiparticles with charge $e$ across the junction, which doubles the Josephson periodicity, $I_c=I_0 \sin(\phi/2)$  \cite{Fu2008}. The doubled periodicity is predicted to lead to the absence of odd integer Shapiro steps in individual junctions, and a SQUID modulation period of 2$\Phi_0$ instead of the usual $\Phi_0$ periodicity, with $\Phi_0=\frac{h}{2e}$ the magnetic flux quantum in superconductivity \cite{Fu2009, Beenakker2011, Badiane2011}.

Experimental efforts have been made to contact superconductors to topologically non-trivial states \cite{Zhang2011, Sacepe2011, Veldhorst12011}. Josephson effects have been observed, and SQUIDs have been reported \cite{Veldhorst22011}. The first signatures of a Majorana fermion, characterized by a zero bias conductance peak, have been observed in superconductor - semiconducting nanowire junctions \cite{Mourik2012}. Nonetheless, so far only $\Phi_0$ periodic dependences have been observed. Relaxation to equilibrium states \cite{Fu2009, Badiane2011}, quantum phase slips \cite{Beenakker2011}, and the large bulk shunt present in contacts with topological insulators so far, may reduce the 2$\Phi_0$ periodicity. 

A key question therefore is how to optimize the Majorana character. Here, we study extrinsic parameters that can be controlled to optimize the  $\sin(\phi/2)$ signal from the Majorana fermion in dc SQUIDs composed of junctions containing both $\sin(\phi)$ and $\sin(\phi/2)$ components in different proportions. Our main observation is that the SQUID parameters $\beta_L$ and $\beta_C$ are important parameters altering the periodicity. Furthermore, a superconducting interferometer will have the periodicity of the component with the smallest periodicity. Nonetheless, even in dc SQUIDs with topologically trivial components the Majorana character strongly influences the dc SQUID characteristics. This study is also of relevance for dc SQUIDs composed of junctions with higher order periodicities, occurring in SNS and SFS systems \cite{Golubov2004}.

\begin{figure}
	\centering 
		\includegraphics[width=0.5\textwidth]{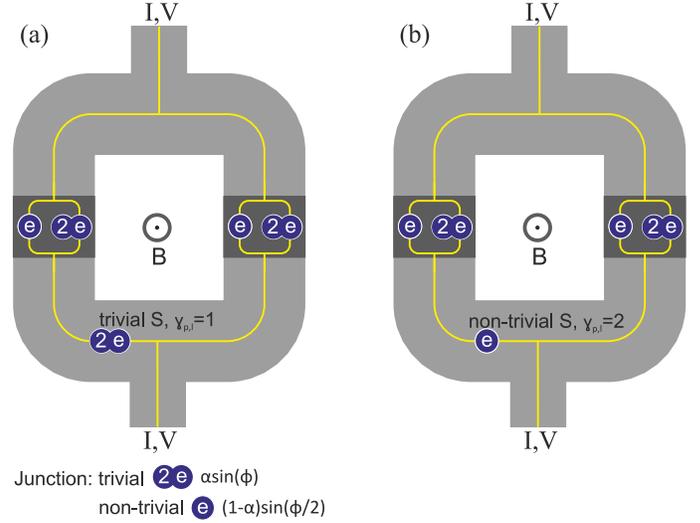}
		\caption{Schematic representation of the considered dc SQUID. The dc SQUID is composed of a superconducting ring interrupted by two Josephson junctions. Charge transport through the Josephson junction is characterized by standard Cooper pair tunneling ($\sin(\phi)$) and single electron tunneling by virtue of the Majorana fermion ($\sin(\phi/2)$). The relative contribution of these two processes is determined by the factor $\alpha$. We analyze two scenarios, in (a) the superconductor is macroscopic and is in the trivial state, $\gamma_{pl}=1$ and in (b) the superconductor is in the topologically non-trivial state, $\gamma_{pl}=2$. Doubled periodicity is only observed when there is a topologically non-trivial state in the entire ring, as in (b). However, the unusual current phase relation changes the dc SQUID characteristics even when the ring is not entirely topologically non-trivial.}
		\label{fig:61}
\end{figure}


\section{Model}
In the next session we introduce fluxoid quantization in superconducting rings composed of topologically trivial and non-trivial parts. Then we will use the fluxoid quantization conditions to determine the critical current in the SQUID under applied magnetic fields and derive the voltage state assuming that the junctions can be described with the resistively and capacitively shunted junction model.

\subsection{Fluxoid quantization in topologically (non)-trivial rings}
The fluxoid quantization in a superconducting loop $\Gamma$ leads to $\gamma_{pl}\Phi_0$ periodicity, with $\gamma_{pl}$ related to the charge carrier $q=\frac{2e}{\gamma_{pl}}$ in the loop. In macroscopic systems, $\gamma_{pl}=1$, but for mesoscopic systems on the order of the superconducting coherence length $\xi$ \cite{Kang2000, Loder2008} and systems including Majorana fermions, $\gamma_{pl}$ can be either 1 or 2 depending on parity conservation \cite{Fu2010}. Integrating the phase of a superconducting loop containing $N$ Josephson junctions results in:

\begin{equation}
\oint \boldnabla \frac{\phi}{\gamma_{pl}} \cdot d\textbf{l}=-\frac{2\pi}{\gamma_{pl} \Phi_0} \int_{\Gamma'} \Lambda \textbf{J}_s\cdot d\textbf{l} -\frac{2\pi}{\gamma_{pl} \Phi_0} \oint \textbf{A}\cdot d\textbf{l}-\sum^{N}_{i=1}\frac{\phi_i}{\gamma_{pj}}
\nonumber
\end{equation}

Here, $\Gamma'$ denotes the contour of the superconducting ring with the Josephson junctions excluded, and $\Lambda$ is a normalization constant for the current. The phase-drop over junction $i$ is given by $\frac{\phi_i}{\gamma_{pj}}$, with $\gamma_{pj}$ connected to the charge carrier $q=\frac{2e}{\gamma_{pj}}$ in the junction. We will consider scenarios where the superconductor is either trivial $\gamma_{pl}=1$ or topologically non-trivial $\gamma_{pl}=2$. Also we will consider the junctions to be trivial $\gamma_{pj}=1$, topologically non-trivial $\gamma_{pj}=2$ or that both charge carrier types are present in the junctions. When the junctions are topologically non-trivial, but the superconductor is macroscopic, quantum phase slips can occur in the superconductor so that $\gamma_{pl}$ can be different from $\gamma_{pj}$. Contour integration over the magnetic vector potential $\textbf{A}$ results in the total flux $\Phi$. Then, in the limit $\textbf{J}_s=0$, assuming thick superconducting leads, fluxoid quantization reduces to: $\frac{1}{2\pi}\sum^{N}_{i=1}\frac{\phi_i}{\gamma_{pj}}+\frac{\Phi}{\gamma_{pl} \Phi_0}=n$. The flux $\Phi$ is the sum of the external flux and the self-flux induced by current flowing through the ring. 

Now we will consider the case of a ring containing two junctions, as depicted in Fig. \ref{fig:61}. We consider the dc SQUID to be symmetric, except for the current phase relationship of the individual junctions. Inclusion of asymmetry (e.g. inductance, critical current or capacitance asymmetry) is easily included. However, inclusion will only lead to asymmetrical SQUID characteristics, and will not change the periodicity. The total flux of the considered system is given by $\Phi=\Phi_e+LI_{c}\chi_1-LI_{c}\chi_2$. Here, $\Phi_e$ is the externally applied flux, $L$ the inductance of a single arm and $I_c$ the critical current of the individual junctions 1 and 2, $I_c=I_{c1}=I_{c2}$.  The factors $\chi_{1,2}$ denote the current dependence on the phase difference of the individual junctions, which we limit to $\chi\chi_0^{-1}=\alpha \sin(\phi)+(1-\alpha) \sin(\phi/2)$, with $\alpha\in[0,1]$ the relative amplitude and $\chi_0$ a normalization factor to have $\max(\chi)=1$. The $\sin(\phi)$ component is the standard Josephson relation, and SNS-junctions are well described by this sinusoidal relation, but it can include higher order components due to $n$ Cooper pairs tunneling \cite{Golubov2004}, described by $I_s(\phi)=\sum^{\infty}_{n=1}I_c^n \sin(n \phi_n)$. Our simplification includes the lowest frequency, which is enough for our conclusions. The $\sin(\phi/2)$ component is due to single electron tunneling by virtue of the Majorana fermion resulting in the 4$\pi$ current phase relationship periodicity.

\subsection{SQUID characteristics in the superconducting and voltage state}
The critical current for an applied external field is obtained by finding the solution of the fluxoid quantization equation with the maximal critical current. The junctions in the voltage state are modeled with the resistively and capacitively shunted junction (RCSJ) model, assuming an ideal Josephson junction shunted by a resistor $R$ and a capacitor $C$: $I=C\frac{dV}{dt}+I_{c} \chi_{1,2}+\frac{V}{R}$. The voltage is related to the time derivative of the phase by $V=\frac{\Phi_0}{2\pi} \frac{d\phi}{dt}$, the same for a topologically trivial and non-trivial ring since we have written the phase as $\frac{\phi}{\gamma_{pj1,2}}$. The RCSJ model leads to the expression $\frac{d^2\phi}{dt^2}+\frac{1}{RC}\frac{d \phi}{dt}+w_{p}^2(\chi_{1,2}-\frac{I}{I_c})=0$, with the plasma frequency $w_{p}=\sqrt{\frac{2\pi}{\Phi_0}\frac{I_{c}}{C}}$. The SQUID parameters are defined as

\begin{eqnarray}
\beta_L & = & \frac{2 \pi L I_{c}}{\Phi_0}, \nonumber \\
\beta_C & = & \frac{2 \pi}{\Phi_0}I_{c}R^2C. \nonumber
\end{eqnarray}

Applying the fluxoid quantization equation, the voltage state can be described by the two differential equations which we have solved numerically

\begin{equation}
\beta_C \frac{d^2\phi_{2,1}}{dt^2}+\frac{d \phi_{2,1}}{dt}+\chi_{2,1}-\frac{1}{2}\frac{I}{I_c}\pm \beta_L^{-1}(\phi_2-\phi_1-2\pi\frac{\Phi_e}{\Phi_0})=0. 
\nonumber
\end{equation}


\section{Results}
In this section we show the SQUID characteristics. We will start analyzing a dc SQUID composed of a topologically trivial ring, and two non-trivial junctions. In this regime there is no doubled fluxoid quantization, however the unusual current phase relationship causes a deviation from standard SQUID characteristics. After that, we will consider the SQUID in the entirely non-trivial regime, where doubled periodicity is observed due to the appearance of the Majorana fermion. Finally, we move to the voltage state and consider both cases in this regime. 

\begin{figure}
	\centering 
		\includegraphics[width=0.45\textwidth]{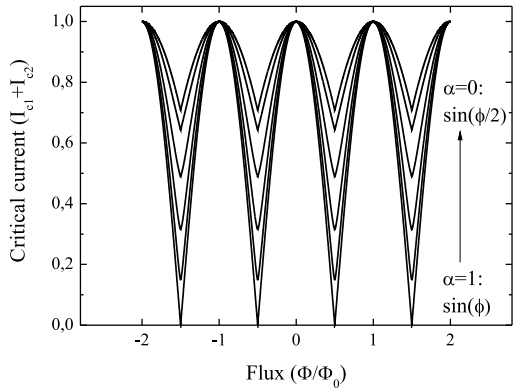}
		\caption{I$\Phi$ characteristics of a dc SQUID composed of a topologically trivial superconducting ring and non-trivial junctions. The current phase relationship of the junctions is shown for $\alpha\sin(\phi)+(1-\alpha)\sin(\phi/2)$ in steps $\delta\alpha=0.2$, $\beta_L=0$. The increase of the $\sin(\phi/2)$ component causes a decrease of the oscillation amplitude without introducing a 2$\Phi_0$ component.}  
		\label{fig:62}
\end{figure}

\subsection{dc SQUIDs composed of trivial and non-trivial elements}
The considered dc SQUID is composed of a topologically trivial ring, and two non-trivial junctions, as shown in Fig. \ref{fig:61}a. Figure \ref{fig:62} shows the critical current dependence of the dc SQUID. In this figure, the non-trivial junctions develop their current phase relationship from pure $\sin(\phi)$ to pure $\sin(\phi/2)$ in steps $\delta\alpha=0.2$. Note that all graphs are calculated for $\beta_L=0$, a situation which for standard SQUIDs leads to a complete critical current modulation. The 2$\Phi_0$ periodicity due to the $\sin(\phi/2)$ component, tends to be completely obscured by the trivial superconducting ring. Quantum phase slips cause the usual $\Phi_0$ periodicity, equivalent to what is calculated by Heck \textit{et al.} \cite{Heck2011} when one of the junctions is topologically trivial. When $\chi_{1}=\sin(\phi/2)$ and $\chi_{2}=\sin(\phi)$ we obtain the result of Fig. \ref{fig:62} for $\alpha$ approximately 0.7, shifted by an additional $\frac{1}{4}\Phi_0$. Instead of $2\Phi_0$ periodicity, the $\sin(\phi/2)$ component influences the modulation depth of the SQUID. There is no appearance of asymmetry as is the case for asymmetric SQUIDs, with different critical currents of the individual junctions \cite{Tesche1977}. The decrease in modulation depth by increasing the $\sin(\phi/2)$ component looks similar to increasing $\beta_L$ in standard SQUIDs. However, this is a parameter that can be controlled externally, and a large $\beta_L$ results in more triangular oscillations. If one junction is topologically trivial, the same effect occurs, combined with a phase shift due to asymmetry between the junctions. Therefore, even in rings including topologically trivial components, a $\sin(\phi/2)$ current phase relationship can be detected, although the effect is more subtle than a 2$\Phi_0$ periodicity. 




\subsection{Topologically non-trivial SQUIDs}
\begin{figure}
	\centering 
		\includegraphics[width=0.5\textwidth]{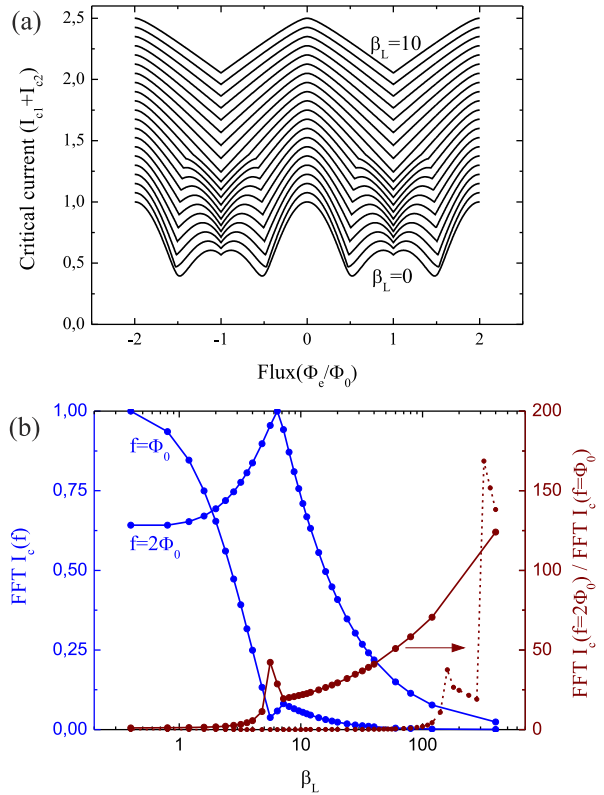}
		\caption{I$\Phi$ characteristics of a dc SQUID composed of a topologically non-trivial superconducting ring and non-trivial junctions. (a) dc SQUID oscillations for two symmetric junctions with equal amplitude $\sin(\phi)$ and $\sin(\phi/2)$ components, $\alpha=0.5$. Increasing $\beta_L$ (in steps $\delta\beta_L=0.5$, and shifted for clarity) promotes the 2$\Phi_0$ period, since the effective screening is smaller for Majorana tunneling than Cooper pair tunneling. (b) (blue) FFT amplitude of the two components as function of $\beta_L$. (red) Evolution of the FFT amplitude ratio as a function of $\beta_L$. The dashed lines represent the ratio when the junctions have only 5$\%$ $\sin(\phi/2)$ component contribution. The 2$\Phi_0$ component can be more than 2 orders of magnitude larger than the $\Phi_0$ component.}  
		\label{fig:63}
\end{figure}

In the case when the ring is completely topologically non-trivial, corresponding to Fig. \ref{fig:61}b, 2$\Phi_0$ periodicity is to be observed. In the limiting case $\beta_L=0$ and $I=I_c \sin(\phi/2)$, the critical current dependence on field can be written as $I=2I_c |\cos(\pi \frac{\Phi}{2\Phi_0})|$, resulting in the 2$\Phi_0$ periodicity. If both $\sin(\phi)$ and $\sin(\phi/2)$ components are present in the junctions, both periodicities are observed, as shown in Fig. \ref{fig:63}a for equal ratios in the junctions. Interestingly, increasing $\beta_L$ results in a larger 2$\Phi_0$ component and a reduced $\Phi_0$ component. In Fig. \ref{fig:63}b the ratio dependence on $\beta_L$ is shown, where the ratio is defined using the frequency amplitude after Fourier transformation. The screening parameter $\beta_L$ is composed of the critical current of the junctions, and the inductance of the ring determined by geometrical factors, but is also dependent on the charge carrier in the ring. Since Majorana tunneling is with charge $e$ instead of $2e$ as is the case for Cooper pair transport, the effective screening is reduced by a factor 2. By optimizing $\beta_L$ using the tunable inductance of the ring, it is therefore possible to dramatically increase the 2$\Phi_0$ component relative to the standard $\Phi_0$-periodic component, ideal for the observation of the Majorana fermion.

\begin{figure*}
	\centering 
		\includegraphics[width=1\textwidth]{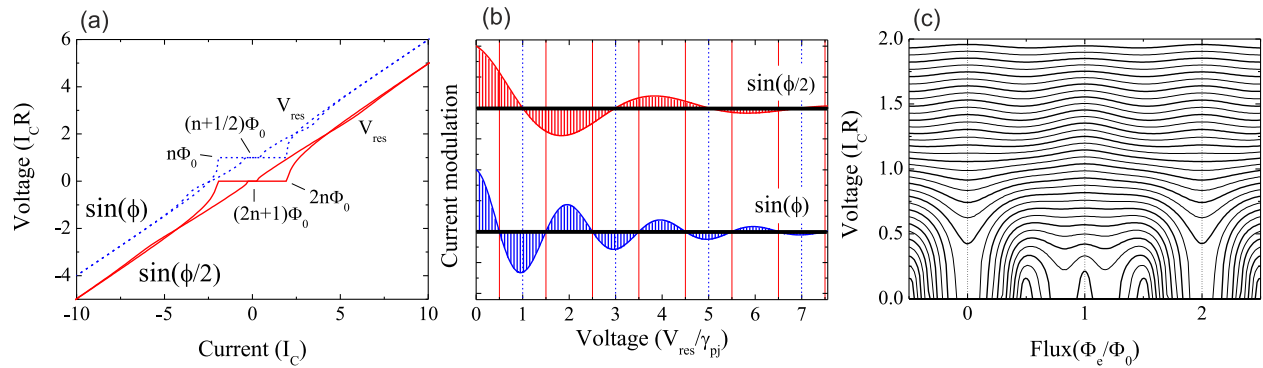}
		\caption{ Topologically non-trivial dc SQUID in the voltage mode. (a) dc SQUID $IV$ characteristics for junctions with $\sin(\phi)$ and $\sin(\phi/2)$ current phase relationships; $\beta_C=1, \beta_L$=1. The $\sin(\phi/2)$ (solid line) component doubles the resonance voltage $V_{res}$ (dashed line) with respect to the standard SQUID. The $IV$ characteristic of the standard SQUID is shifted for clarity. (b) Damping and resonance in the SQUID. The $\sin(\phi/2)$ component oscillates with half the frequency, so that the $\sin(\phi)$ and $\sin(\phi/2)$ components have their minima and maxima at different voltages (solid red are maxima for the $\sin(\phi/2)$ component and dashed blue maxima for the $\sin(\phi)$ component. (c) $V\phi$ characteristics of a dc SQUID with equal $\sin(\phi)$ and $\sin(\phi/2)$ components in the junctions, and $\beta_C=1, \beta_L=1$. The SQUID shows both $\Phi_0$ and $2\Phi_0$ periodicity, depending on the voltage close to $V_{res,\phi}$ or $V_{res,\phi/2}$, which is controlled by the bias current.}  
		\label{fig:64}
\end{figure*}

\subsection{The voltage state}
When the SQUID is operated in the voltage mode, both $\Phi_0$ and 2$\Phi_0$ periodicity can be observed, even if the ring includes trivial components. The relative amplitude depends on the voltage, controlled by the bias current. Figure \ref{fig:64}a shows the $IV$ characteristics for a dc SQUID with either pure $\sin(\phi)$ or $\sin(\phi/2)$ components. The current modulation for high voltages is inverted with respect to the modulation for small voltages. This is the result of the nonlinear interaction of the ac Josephson current with the resonant circuit formed by the loop inductance L and the junction capacitance $C$. This resonance voltage is given by $V_{res}=\gamma_{pj} \sqrt{\frac{2}{\beta_C \beta_L}}I_c R$. As a result, the $\sin(\phi)$ and $\sin(\phi/2)$ components cause oscillations with different frequency as a function of voltage, see Fig. \ref{fig:64}b. Consequently, the relative amplitude of the 2 and 4$\pi$ periodicity of the $V(\phi)$ characteristics, shown in Fig. \ref{fig:64}c for $\beta_C=1$, and $\beta_L=1$ depend on the resonance voltages $V_{res,\phi}$ and $V_{res,\phi/2}$, which is controlled by the bias current. The resonance voltage is independent of the loop parity, but for $\gamma_{pl}=1$ the oscillation amplitude is reduced by quantum phase slips. This is similar to the cause of the incomplete critical current modulation shown in Fig. \ref{fig:62}.

The damping of the current modulation with increasing bias voltage is characterized by $\zeta=\sqrt{\frac{2\beta_C}{\beta_L}}$, independent on the current phase periodicity, and SQUIDs with $\sin(\phi/2)$ component junctions have therefore a smaller current modulation where the modulation is inverted, compared to standard SQUIDs. Choosing a large $\beta_L$ will decrease the damping term and increase the amplitude of the voltage resonances. A small $\beta_C$ will reduce the damping term but increase the resonance voltage.

\section{Applications to topologically non-trivial systems}
We now discuss the implications of our proposals. The possible realization of a new emergent particle in condensed matter physics together with the potential for quantum computation, boosted the search for superconducting systems hosting the Majorana fermion. Superconductor - semiconductor structures in the presence of strong spin orbit coupling and Zeeman fields are currently quite successful and signatures of Majorana fermions have been observed characterized by zero bias conductance peaks \cite{Mourik2012}. In superconductor - topological insulator junctions, Josephson supercurrents have also been observed \cite{Zhang2011, Sacepe2011, Veldhorst12011, Qu2012}. In the case of topological insulator systems, bulk shunting likely introduces $\sin(\phi)$ terms in the current phase relation. Quantum phase slips \cite{Beenakker2011} and quasiparticle poisoning \cite{Fu2009} will be relevant for all proposals, relaxing the system to $\sin(\phi)$ periodicity. Therefore, increasing the $\sin(\phi/2)$ component is important for all these proposals.

The parameter $\beta_L$ is easily tunable. This SQUID parameter depends on the inductance, determined by the SQUID geometry, and independent of the individual junctions. The parameter $\beta_C$ is a junction parameter and therefore more difficult to tune. However, varying the length and width of the junctions and controlling the interface transparency tune $\beta_C$.

\section{Conclusions}
In conclusion, we have studied dc SQUIDs containing two topologically non-trivial barriers. The 2$\Phi_0$ periodicity stemming from the Majorana fermion can only be detected in non-trivial dc SQUIDs. However, even loops containing topologically trivial elements are influenced by the presence of junctions with $\sin(\phi/2)$ components. This is observed both in the critical current modulation by flux and the resonance voltage. The SQUID parameters can be used to increase the relative component; increasing $\beta_L$ is found to largely increase the component with the largest periodicity. The $V(\phi)$ relation is altered when both components are present, and both components can be maximized at different bias currents, determined by the resonance voltage. In recently fabricated S-TI-S junctions, $\beta_C$ is usually low due to bulk shunting. This increases the resonance voltage to a regime where the damping is higher, which complicated observing a clear difference between $\Phi_0$ and $2\Phi_0$ periodicities. Decreasing this bulk shunt would therefore simplify the 2$\Phi_0$ periodicity observation. Nonetheless, the strong effect of $\beta_L$; large $\beta_L$ increases the 2$\Phi_0$ component over 100 times, allows  the detection of the Majorana fermion, under the right intrinsic circumstances (e.g. no relaxation to equilibrium), in dc SQUIDs composed of present day S-TI-S junctions. This result is also of relevance to devices where the Majorana character is induced via other means, such as in semiconducting nanowires with strong Rashba spin orbit coupling. Tuning $\beta_L$ in combination with ac measurements to prevent relaxation paves the way to the observation of exotic properties of the Majorana fermion.

We acknowledge A. Andreski, A.A. Golubov, and M. Snelder for useful discussions. This work is supported by the Netherlands Organization for Scientific Research (NWO) through VIDI and VICI grants, and the Dutch FOM foundation.


\begin{thebibliography}{1}
\bibitem{Fu2008} L. Fu and C.L. Kane Phys. Rev. Lett. \textbf{100}, 096407 (2008).   
\bibitem{Tanaka2009} Y. Tanaka, T. Yokoyama, and N. Nagaosa, Phys. Rev. Lett. \textbf{103}, 107002 (2009). 

\bibitem{Zhang2006} B.A, Bernevig, T.L. Hughes, and S.C. Zhang Science \textbf{314}, 1757 (2006).
\bibitem{Fu2007} L. Fu, C.L. Kane, and E.J. Mele Phys. Rev. Lett. \textbf{98}, 106803 (2007). 
\bibitem{Hsieh2008} D. Hsieh, D. Qian, L. Wray, Y. Xia, Y.S. Hor, R.J. Cava, and M.Z. Hasan, Nature \textbf{452}, 970 (2008). 
\bibitem{Zhang2009N} H. Zhang, C.X. Liu, X.L. Qi, X. Dai, Z. Fang, and S.C. Zhang, Nature Phys. \textbf{5}, 438 (2009). 
\bibitem{Qi2009} X.L. Qi, L. Rundong, J. Zang, and S.C. Zhang, Science \textbf{323}, 1184 (2009). 
\bibitem{Chen2009} Y.L. Chen, J.G. Analytis, J.H. Chu, Z.K. Liu, S.K. Mo, X.L. Qi, H.J. Zhang, D.H. Lu, X. Dai, Z. Fang, S.C. Zhang, I.R. Fisher, Z. Hussain, and Z.X. Shen, Science \textbf{325}, 178 (2009). 
\bibitem{Hsieh2009N} D. Hsieh, Y. Xia, D. Qian, L. Wray, J.H. Dil, F. Meier, J. Osterwalder, L. Patthey, J.G. Checkelsky, N.P. Ong, A.V. Fedorov, H. Lin, A. Bansil, D. Grauer, Y.S. Hor, R.J. Cava, and M.Z. Hasan, Nature \textbf{460}, 1101 (2008). 
\bibitem{Peng2009} H. Peng, K. Lai, D. Kong, S. Meister, Y. Chen, X.L. Qi, S.C. Zhang, Z.X. Shen, and Y. Cui,  Nature Mat. \textbf{9}, 225 (2010). 
\bibitem{Zhang2009P} T. Zhang, P. Cheng, X. Chen, J.F. Jia, X. Ma, K. He, L. Wang, H. Zhang, X. Dai, Z. Fang, X. Xie, and Q.K. Xue, Phys. Rev. Lett. \textbf{103}, 266803 (2009). 
\bibitem{Cheng2010} P. Cheng, C. Song, T. Zhang, Y. Zhang, Y. Wang, J.F. Jia, J. Wang, Y. Wang, B.F. Zhu, X. Chen, X. Ma, K. He, L. Wang, X. Dai, Z. Fang, X. Xie, X.L. Qi, C.X. Liu, S.C. Zhang, and Q.K. Xue, Phys. Rev. Lett. \textbf{105}, 076801 (2010). 

\bibitem{Sau2010} J.D. Sau, R.M. Lutchyn, S. Tewari, S. Das Sarma, Phys. Rev. Lett. \textbf{104}, 040502 (2010).
\bibitem{Alicea2010} J. Alicea, Phys. Rev. B \textbf{81}, 125318 (2010).

\bibitem{Majorana1937} E. Majorana, Nuovo Cimento \textbf{14}, 171 (1937). 

\bibitem{Fu2009} L. Fu and C.L. Kane, Phys. Rev. B \textbf{79}, 161408(R) (2009).
\bibitem{Beenakker2011} C.W.J. Beenakker, Arxiv:1112.1950v2 (2012).
\bibitem{Badiane2011} D.M. Badiane, M. Houzet and J.S. Meyer, Phys. Rev. Lett \textbf{107}, 177002 (2011).


\bibitem{Zhang2011} D. Zhang, J. Wang, A.M. DaSilva, J.S. Lee, H.R. Gutierrez, M.H.W. Chan, J. Jain, and N. Samarth, Phys. Rev. B \textbf{84}, 165120 (2011).  
\bibitem{Sacepe2011} B. Sac$\acute{\textrm{e}}$p$\acute{\textrm{e}}$, J.B. Oostinga, J.L. Li, A. Ubaldini, N.J.G. Couto, E. Giannini, and A.F. Morpurgo, Nature Comm. \textbf{2}, 575 (2011).  
\bibitem{Veldhorst12011} M. Veldhorst, M. Snelder, M. Hoek, T. Gang, X.L. Wang, V.K. Guduru, U. Zeitler, W.G. v.d.Wiel, A.A. Golubov, H. Hilgenkamp, and A. Brinkman, Nature Mat. \textbf{11}, 417 (2012). 
\bibitem{Veldhorst22011} M. Veldhorst, C.G. Molenaar, X.L. Wang, H. Hilgenkamp, and A. Brinkman, Appl. Phys. Lett. \textbf{100}, 072602 (2012). 
\bibitem{Mourik2012} V. Mourik, K. Zuo, S.M. Frolov, S.R. Plissard, E.P.A.M. Bakkers, and L.P. Kouwenhoven, Science \textbf{336}, 1003 (2012).










\bibitem{Golubov2004} A.A. Golubov, M.Y. Kupriyanov, and E. Il'ichev, Rev. Mod. Phys. \textbf{76}, 411 (2004).
\bibitem{Kang2000} K. Kang, Eur. Phys. Lett. \textbf{51}, 2 (2000).
\bibitem{Loder2008} F. Loder, A.P. Kampf, and T. Kopp, Phys. Rev. B \textbf{78}, 174526 (2008).
\bibitem{Fu2010} L. Fu, Phys. Rev. Lett. \textbf{104}, 056402 (2010).


\bibitem{Heck2011} B. van Heck, F. Hassler, A.R. Akhmerov, and C.W.J. Beenakker, Phys. Rev. B \textbf{84}, 180502(R) (2011).

\bibitem{Tesche1977} C.D. Tesche and J.C Clarke, J. Low. Temp. Phys. \textbf{29}, 301 (1977).
\bibitem{Qu2012} F. Qu, F. Yang, J. Shen, Y. Ding, J. Chen, Z. Ji, G. Liu, J. Fan, X. Jing, C. Yang, and L. Lu, Sci. Rep. \textbf{2} 339.
\end{thebibliography}
\end{document}